\documentclass[sigconf]{acmart}

\renewcommand\footnotetextcopyrightpermission[1]{}
\setcopyright{none}
\usepackage{multirow}

\AtBeginDocument{%
  \providecommand\BibTeX{{%
    \normalfont B\kern-0.5em{\scshape i\kern-0.25em b}\kern-0.8em\TeX}}}
\begin{document}

\title{Semantic Search Evaluation}

\author{Chujie Zheng}
\authornote{Corresponding author, email: razheng@linkedin.com.}
\affiliation{%
\institution{LinkedIn}
\city{Sunnyvale}
\state{California}
\country{USA}
}

\author{Jeffrey Wang}
\affiliation{%
\institution{LinkedIn}
\city{Sunnyvale}
\state{California}
\country{USA}
}

\author{Shuqian Albee Zhang}
\affiliation{%
\institution{LinkedIn}
\city{Sunnyvale}
\state{California}
\country{USA}
}

\author{Anand Kishore}
\affiliation{%
\institution{LinkedIn}
\city{Sunnyvale}
\state{California}
\country{USA}
}

\author{Siddharth Singh}
\authornote{Work done in LinkedIn.}
\affiliation{%
\institution{Walmart}
\city{Sunnyvale}
\state{California}
\country{USA}
}

\renewcommand{\shortauthors}{Zheng and Wang, et al.}

\begin{abstract}
We propose a novel method for evaluating the performance of a content search system that measures the semantic match between a query and the results returned by the search system. We introduce a metric called "on-topic rate" to measure the percentage of results that are relevant to the query. To achieve this, we design a pipeline that defines a golden query set, retrieves the top K results for each query, and sends calls to GPT 3.5 with formulated prompts. Our semantic evaluation pipeline helps identify common failure patterns and goals against the metric for relevance improvements.
\end{abstract}


\maketitle
\pagestyle{plain}
\section{Introduction}
LinkedIn search has made significant progress over the years by incorporating semantic matching capabilities. Semantic matching capability helps members find knowledge more easily by delivering results that are conceptually related to their search query, even if they do not contain the exact keywords used in the query. As we continue to improve our system, we strive to overcome the complex quality challenges that arise:
\begin{itemize}
    \item Indirect measurement: While we have existing engagement metrics from online experiments, they do not necessarily capture the quality of our search results. When member feedback comes in, there is an added overhead in determining if the system is operating as expected. 
    \item Not operationalized: Search patterns and expectations change over time, so we need an automated way to continuously measure quality.
\end{itemize}

To address these gaps, we present a semantic evaluation pipeline that leverages Generative AI (GAI) to evaluate quality. Our main contributions and business impacts are:
\begin{itemize}
    \item We propose the metric “on-topic rate” to measure the percentage of content search results that are relevant to the intended topic. This metric serves as a tool for evaluating the performance of the content search model, and can also help identify common failure patterns. By setting goals based on this metric, we can work towards improving the relevance of search results.
    \item We present a novel semantic evaluation pipeline for search engine offline evaluation. This framework helps translate user problems into technical patterns that can be operationalized and plays a critical role for search engine offline evaluation.
    \item We present two approaches to evaluate the Generative AI evaluation output, including conducting human evaluation and preparing a validation set to ensure the quality.
\end{itemize}

\section{Related Work}
Assessing the quality of search engine results has become a challenging task, particularly in determining which ranking models perform best under specific business metrics \cite{carmel2020multi, karmaker2017application}. The most reliable method for evaluating model performance is through online A/B testing, but this approach has limitations \cite{gupta2019top, johari2017peeking}. Firstly, due to the high cost of traffic, only a limited number of models can be compared within a given timeframe, as it requires a significant amount of user feedback to draw statistically significant conclusions. Secondly, there is a risk of negatively impacting the user search experience if the model performs poorly. Consequently, offline evaluation is commonly used to select candidates for online experiments.

Existing work has proposed a variety of techniques for search engine offline evaluation, including relevance, reliability, timeliness, diversity and fairness \cite{chen2023practice}. Commonly we evaluate the quality of search by using the evaluation metrics focused on relevance, such as Mean Average Precision (MAP) and normalized Discounted cumulative gain (nDCG). These types of metrics simulate how users interact with the search engine result page (SERP) \cite{chen2017meta, sanderson2010test}. 
Nonetheless, a prevalent drawback of offline evaluation lies in its dependence on historical data, and these metrics may not directly measure the quality of search engine, which can result in imperfect correlations with users' search experiences \cite{Wang2023, al2007relationship, huffman2007well, mao2016does}. 

Recent advancements in semantic search and generative AI have significantly transformed the landscape of search systems, particularly with the integration of large language models (LLMs) in ranking and recommendation tasks \cite{zhuang2024setwise, wan2024larr, li2024corpuslm}. LLMs are also explored for relevance judgments and evaluation in information retrieval \cite{rahmani2024report}. The literature has shown that LLMs can closely replicate human judgements through prompt strategies \cite{rahmani2024llmjudge} and highlights the growing role of LLMs in automated evaluations \cite{rahmani2024llm4eval}. 

\section{On-Topic Rate}
In this section, we formulate our task and introduce our proposed metric - On-Topic Rate (OTR). This metric is proposed as the direct measurement to evaluate if the retrieved document is primarily about the query.

\subsection{Task Formulation}
Given a member query $q$ as input, the search engine returns a list of documents $D=(d_1, d_2, ..., d_n)$. Since we focus on content search, here each document corresponds to a post or article on LinkedIn. 

\subsection{Computation}
On-topic rate is a metric that measures the relevance of search results to the user's query. 
This metric helps evaluate how well the search engine is able to understand the query's intent and provide relevant results. A high on-topic rate indicates that content search is performing well and providing useful results to our members, while a low on-topic rate suggests that improvements may be needed to better understand the user's search intent. We define OTR for <query, doc> pair as the following: 
\begin{equation}
    On Topic Rate (q, d_i)=1 \text{ if the pair is relevant, otherwise 0}
\end{equation}

\subsubsection{OTR@K}
We measure the on-topic rate for the ML model by selecting the top $K$ returned documents for each query. We define OTR@K as the total number of query-document pairs that are relevant divided by the total number of returned documents.
\begin{equation}
    OTR@K = \frac{\sum_{i=1}^{K}OTR(q, d_i)}{K}
\end{equation}




\begin{figure*}
\centering
\includegraphics[width=1\textwidth]{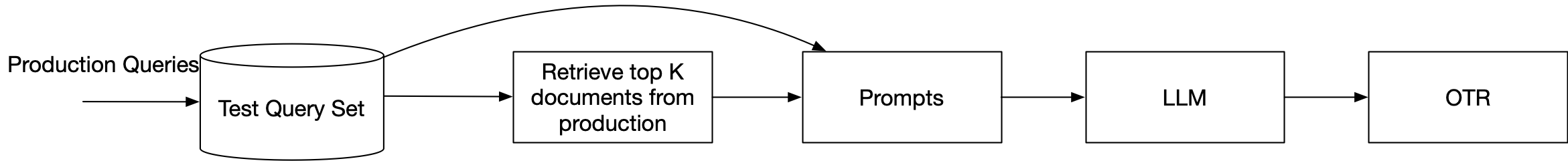}
\caption{Semantic evaluation pipeline}
\label{fig::otr-pipeline}
\end{figure*}

\section{Semantic Evaluation Setup}
Within this section, we detail the process of establishing semantic evaluation and harnessing Generative AI (GAI) to discern the relevance between queries and documents. Figure \ref{fig::otr-pipeline} has illustrated the overall pipeline: our methodology begins with the creation of a test query set, followed by the formulation of prompts provided to the Large Language Model (LLM) and subsequent calculation of OTR based on the GAI-generated outputs. 

\subsection{Create Query Set}
We construct the query set used for evaluation by leveraging different resources to comprehensively cover the relevant topics or areas of interest. 

\subsubsection{Golden Set}
The golden set serves as a stable and uniform standard for assessing and benchmarking queries. It includes the aspirational queries and the top queries from production. We include the following types of query into the golden set:
\begin{itemize}
    \item Top queries: We incorporate common queries, which can be seen as the most popular and indicative keywords that members are searching for.
    \item Topical queries: Topical queries encompass search inquiries or questions that pertain to specific subjects or topics. We incorporate these into our golden set because they pose a greater evaluation challenge, often being lengthier and more intricate in their intent. Providing high-quality results for topical queries can enhance our members' ability to access valuable knowledge on LinkedIn.
\end{itemize}



\begin{table}[htbp]
\centering
\begin{tabular}{l|p{5cm}}
\toprule
 & \textbf{Example Queries} \\
 \hline
Top queries & covid-19, resume, microsoft excel, we're hiring, work from home\\
\hline
Topical queries & how to create a personal brand, how to stand out in a competitive job market, how do I negatiate my salary\\
\toprule
\end{tabular}
\caption{Example golden set queries from different sources}
\label{table::example-golden-set-query}
\end{table}

\subsubsection{Open Set}
The Open Set is a dynamically changing set of queries used for evaluation. The source of the open set includes:
\begin{itemize}
    \item trending queries and newsy queries from production
    \item some random queries from production, to add diversity
\end{itemize}

Table \ref{table::example-query} are some example queries from the golden set and open set used in production for evaluation.

\begin{table}[htbp]
\centering
\begin{tabular}{l|p{6cm}}
\toprule
 & \textbf{Example Queries} \\
 \hline
Golden Set & improve workplace communication, remote team best practices, how do I get promoted\\
\hline
Open Set & fed raises rates,leadership first, barbie, women ai study \\
\toprule
\end{tabular}
    \caption{Example queries in golden set and open set}
    \label{table::example-query}
\end{table}

\subsection{Get search results for query set}
We gather the following information for each query within the specified set: the top $K=10$ documents from production, along with post textual information and any mentioned article information (including title and article body, if applicable).

\subsection{Formulate the prompt}
We construct the prompt for GPT 3.5 to collect feedback from the LLM. Using the inputs collected from previous steps, we define the prompt to encompass three perspectives: 
\begin{enumerate}
    \item the definition of On-Topic rate
    \item detailed guidance for decision making
    \item query
    \item post, including all the text information from the posts, including the commentary and any re-shared posts/articles
\end{enumerate}

\subsubsection{Metric Definition}
We begin by defining a metric and then task LLM with making decisions based on our specific requirements. We keep iterating our prompts to improve the accuracy of the decision. Based on our experiences from production and our prompting practices, we have learnt that providing very precise definition for the request and adding examples with provided reasons can improve the performance significantly.

For example, for our task, we tested two prompts in production for our use case:
\begin{itemize}
    \item \textbf{Prompt A}: Given the post below, is the post strongly relevant to the query?
    \item \textbf{Prompt B}: Given the post below, is the post primarily about query or strongly relevant to the query?
\end{itemize}
Despite the subtle differences between the two prompts, Prompt B outperforms Prompt A significantly. This is because Prompt B specifically directs attention to the main subject of both the query and the post. As a result, it reduces the false positive that might occur due to keyword matching. We have included some examples to show the impacts in Appendix \ref{appendix:examples-prompt-a-and-b}.

We discover that incorporating well-detailed examples alongside the reasons behind the decision can enhance performance and address various problems identified from observed patterns of failure. With no examples in the prompt the results did not align with human judgment. By summarizing the failure pattern, during our prompt iteration, we determine that incorporating examples with explanations of the judgments can boost the generative AI's capacity for better decision making.

\subsubsection{Guidance}
The guidance offers comprehensive instructions for training LLM to make decisions that align with our expectations. This may involve outlining decision criteria and specific requirements, as well as providing examples of corner cases to illustrate decision-making processes. 

Here is an example of defining guidance for OTR:
\begin{enumerate}
    \item The on-topic decision should not only consider the keyword match between query and post. It should reflect the semantic matching between query intention and the post details.
    \item The post information is primarily relevant to the user query. 
\end{enumerate}


\subsection{Compute OTR Metrics}
From the GAI output, we have three types of information: 
\begin{itemize}
    \item Binary decision: It directly corresponds to if the retrieved post is primarily on-topic of the query. 
    \item Relevance score: It is the score related to the binary decision. The score is in the range of 0 and 1. The relevance score aims to measure the semantic relevance between query and post. The decision should keep the consistency between the relevance score.
    \item Decision reason: It explains the reason for the binary decision and relevance score. This field is not used for OTR calculation but it is very important to help us iterate the prompt, since it explains the consideration for the GAI decision.
\end{itemize}

Table \ref{table::example-gai-output} presents an instance from the pipeline where a post mentioning an article introducing tips for self-promotion was retrieved for the query "promotion tips". Although this example may be classified as on-topic through keyword matching, it does not accurately reflect the query intention. To address this issue, semantic evaluation enables the identification of the primary topic of the post. Based on the decision reason, it can be concluded that the prompt provided instructions for measuring semantic relevance, and the LLM was able to identify that although the keyword matched the query, the post itself did not address the query intention, resulting in a low relevance score and an off-topic binary decision.

\begin{table}[htbp]
\centering
\begin{tabular}{l|p{5.5cm}}
\toprule
Query & promotion tips \\
\hline
Post & Here are 13 tips to get you over that mental hurdle. \#speakup \#tips \#leadership\\
\hline
Binary decision & 0\\
\hline
Relevance score & 0.4\\
\hline
Decision Reason & The post is about tips for self-promotion and personal branding, and does not directly address the query of "promotion tips". The mention of "tips" in the hashtags is not specific enough to make a strong connection to the query.\\
\toprule
\end{tabular}
    \caption{Example output from semantic evaluation pipeline.}
    \label{table::example-gai-output}
\end{table}

Our approach consider (query, document) as on-topic only if the binary decision is 1 and the relevance score exceeds a certain threshold (we use 0.5 in production after analyzing the distribution of relevance score). This ensures that we only consider pairs that the GAI has a strong confidence in, and disregard those that fall below the threshold. By taking this rule, we compute $OTR@K$ and nDCG as the final output.

\section{Experiment}
\subsection{Human Evaluation on Generated Output}
We employ human evaluation as a metric to ascertain the alignment between decisions made by the proposed semantic evaluation pipeline and expert human annotators, aiming to verify the reliability and consistency of the proposed pipeline in delivering high-quality results.

Our human evaluation team comprises 10 colleagues, and all team members possess substantial experience in content search to ensure the credibility of their decisions. To maintain the quality of the annotation process, annotators underwent a rigorous qualification process. This process included familiarizing themselves with annotation guidelines and assessing 20 representative query-response pairs. Following this, we conducted individual assessments of the annotations submitted and organized group discussions to address any confusions or uncertainties regarding the task. Each annotator was given the task of annotating 50 pairs of queries and documents. Both annotators and the GAI are providing the same information: query, post commentary, and re-shared posts/articles.

To gauge the level of agreement among annotators, we collected annotations for a specific set of randomly selected query-response pairs, which yielded a high degree of agreement among the annotators.

Evaluating a query-documents pair requires our annotators to complete a three-step evaluation:
\begin{enumerate}
    \item Evaluate if the query and the post is relevant
    \item Provide reasons regarding the judgement if it is considered as irrelevant
\end{enumerate}

We utilize the (query, document) data from the annotation result as input for the proposed semantic evaluation process and then compare the output of the pipeline with human evaluation to assess its consistency. By comparing the results, we report 81.72\% consistency with the GAI decision.
\subsection{Performance on Validation Set}
To ensure the quality of our prompt, the team has compiled a validation set comprising various query types and corresponding post pairs, serving as the baseline for assessing the prompt's effectiveness. We select representative queries from production that are frequently searched by members, including:
\begin{itemize}
    \item company name queries
    \item title queries, like data engineer, product manager
    \item skill queries, like finance, customer services, marketing
    \item newsy queries, like february jobs report
    \item other top queries, for example: work from home, open to work
\end{itemize}

Total our validation set includes 60 queries, each paired with 10 related posts, resulting in 600 query-post pairs. Our team assesses these pairs and makes a binary determination regarding their topical relevance. This judgment is confirmed by at least three team members. We use this validation set as our standard of truth for gauging the effectiveness of our prompts. Furthermore, we utilize this validation set to identify common failure patterns to keep iterating our prompts. The current prompt used in production achieves 94.5\% accuracy on the validation set.

\section{How do we use semantic evaluation to improve the product?}
We have adopted the semantic evaluation pipeline as our offline benchmark for experiments in LinkedIn content search. This serves as the foundation for offline evaluation, measuring whether the trained ML model has correctly captured the query intention. The pipeline is designed to monitor the performance of the served content search model, with a weekly update of the open query set and pipeline run to ensure that the calculated OTR falls within the desired range. We also use this tool to evaluate new trained ML models offline, comparing them to the baseline to quickly test for improvements and provided feedback for iteration.

In addition, we explore the decision reasons to identify growth opportunities. We collect cases identified as off-topic from the pipeline and identify performance gaps that we can improve.

\section{Conclusion}
In this work, we introduce a semantic evaluation pipeline for search engine offline evaluation. Our proposed metric, on-topic rate, measures the relevance of search results to the user's query. We also outline the construction of prompts and calculation of the OTR, and demonstrate the high consistency between human evaluation and pipeline output. With this semantic evaluation framework, we are able to directly measure the quality of post-search results and gain a better understanding of our search engine's performance.

\section*{Acknowledgments}
The authors would like to thank Rupesh Gupta, Raghavan Muthuregunathan, Dhruv Saksena, Abhi Lad, Alice Xiong, Rashmi Jain for their support of the work. We also acknowledge and thank the contributions and feedback from our colleagues in LinkedIn Content Search AI team: Xin Yang, Madhumitha Mohan, Ali Hooshmand, Justin Zhang and Sarang Metkar.

\bibliographystyle{ACM-Reference-Format}
\bibliography{literature}

\appendix
\section{Examples to demonstrate the performance improvement between Prompt A and Prompt B}
\label{appendix:examples-prompt-a-and-b}
During our experiments, we have noticed the importance of formatting the request into very precise guidance that can help GAI understand the request and leverage it for decision making. For example, for two prompts mentioned earlier, given the same information for query and posts, we are seeing Prompt B outperforms Prompt A by significantly reducing the false positives caused by keyword matching.
\begin{itemize}
    \item \textbf{Prompt A}: Given the post below, is the post strongly relevant to the query?
    \item \textbf{Prompt B}: Given the post below, is the post primarily about query or strongly relevant to the query?
\end{itemize}

Here are some examples that Prompt A and Prompt B are making different decisions in Table \ref{tab:examples-from-prompt-a-and-b}.
\begin{table*}[h]
\centering
\begin{tabular}{c|p{8cm}|c|c}
\toprule
\textbf{Query} &  \multicolumn{1}{c}{\textbf{Post}} & \textbf{Prompt A} & \textbf{Prompt B} \\
\hline
react native  &  Hey LinkedIn community,

In my recent article, "The Rising of New Gen-GPT and Socializing as an Adult," I dove into some exciting topics that I want to share with you in a nutshell:

Projects in Progress: I've been working on various projects, including exploring the Spotify API for a dynamic playlist-based website and a new React Native project. The latter aims to address real-world needs, from allergen tracking to task delegation tools and a notes app integrated with Slack to help in my current position.

AI Evolution: The latest Gen-GPT update is a game-changer

Balancing Act: I shared my personal reflections on the importance of work-life balance. While pursuing my career, I realized the value of reconnecting with friends, especially those from my high school days, as a source of relaxation and genuine joy.

These are just the highlights. If you want to dive deeper into these topics, check out the full article linked in this post.

Your feedback and insights are always appreciated. Let's keep the conversation going!

I write them weekly it is something will not want to miss. & On Topic & Off Topic \\
\hline
manager  &  Twin Transformation is currently THE topic of sustainability managers, but also of digitalisers. Brigitte Falk was a pioneer in this field 20 years ago. So her interview (Digitization as a bridge to sustainable business) is hopefully an inspiration and encouragement for many as a look back, but above all as a look forward! & On Topic & Off Topic \\
\hline
best practices for managing remote teams  & Excited to invite you to my upcoming workshop at \#ReactIndia 2023 with my co-speaker Lokesh Yadav on "Building a Web Performance Culture: Empowering Large-Scale Teams to Deliver Lightning-Fast User Experiences"!

In today's digital world, web performance plays a crucial role in user experience, business success, and stakeholder satisfaction. Slow-loading websites can significantly impact user engagement, conversion rates, and revenue. That's why it's essential for large-scale teams to establish a web performance culture and empower their teams to deliver exceptional user experiences.

Join me for this workshop where we'll dive deep into the key aspects of building a web performance culture within larger teams. We'll explore the significance of performance for you and your stakeholders, and learn how to differentiate between noise and reality by implementing the right tooling for Single-Page Applications (SPA) and Multi-Page Applications (MPA).

Attendees will gain valuable insights into improving performance metrics and discover innovative approaches using Real User Monitoring (RUM) and Lab data to track and monitor performance enhancements without incurring additional costs. We'll also cover practical strategies, best practices, and real-life examples, including tackling performance challenges with React 18 Hydration.

Whether you're a developer, manager, or part of a large-scale team, this workshop will help you to build new perspectives towards web performance and empower you to deliver lightning-fast user experiences.  & Off Topic & On Topic \\
\toprule
\end{tabular}
\caption{Examples from Prompt A and Prompt B that are making different decisions.}
\label{tab:examples-from-prompt-a-and-b}
\end{table*}

\end{document}